\documentclass[prl,twocolumn,floatfix,a4paper,nofootinbib,amssymb,amsmath,showpacs,superscriptaddress]{revtex4}
\usepackage{graphicx}
\usepackage{bm}
\usepackage{eucal}
\usepackage{mathrsfs}

\newcommand {\be} {\begin{equation}}
\newcommand {\ee} {\end{equation}}
\newcommand {\Be}{\begin{eqnarray*}}
\newcommand {\Ee} {\end{eqnarray*}}
\newcommand {\bey} {\begin{eqnarray}}
\newcommand {\eey} {\end{eqnarray}}
\newcommand{\bit}{\begin{itemize}}      
\newcommand{\eit}{\end{itemize}}
\newcommand{\bfl}{\begin{flusleft}}
\newcommand{\efl}{\end{flusleft}}
\newcommand{\bfr}{\begin{flushright}}
\newcommand{\bc}{\begin{center}}
\newcommand{\ec}{\end{center}}
\newcommand{\ben}{\begin{enumerate}}    
\newcommand{\een}{\end{enumerate}}

\newcommand{\comment}[1]{}

\newcommand{\E}{\mathrm{e}}

\newcommand{\D}{\mathrm{d}}

\newcommand{\average}[1]{\left\langle{#1}\right\rangle}

\begin{document} 

\title{Reconstructing the free energy landscape of a mechanically unfolded model protein}

\author{Alberto Imparato}
\email{alberto.imparato@polito.it}
\affiliation{Dipartimento di Fisica, INFN Sezione di Torino,
CNISM-Sezione di Torino\\ Politecnico di Torino, Corso Duca degli
Abruzzi 24, 10129 Torino, Italy}
\author{Stefano Luccioli}
\affiliation{Istituto dei Sistemi Complessi, CNR, via Madonna del Piano 10, 50019 Sesto Fiorentino, Italy}
\affiliation{INFN Sezione di Firenze, via Sansone, 1 - 50019 Sesto Fiorentino, Italy}
\author{Alessandro Torcini}
\affiliation{Istituto dei Sistemi Complessi, CNR, via Madonna del Piano 10, 50019 Sesto Fiorentino, Italy}
\affiliation{INFN Sezione di Firenze, via Sansone, 1 - 50019 Sesto Fiorentino, Italy}

\begin{abstract}
The equilibrium free energy landscape of an off-lattice model protein as a
function of an internal (reaction) coordinate is reconstructed from
out-of-equilibrium mechanical unfolding manipulations. This task is 
accomplished via two independent methods: by employing an extended version 
of the Jarzynski equality (EJE) and the protein inherent structures (ISs).
In a range of temperatures around the ``folding transition'' 
we find a good quantitative agreement between the free energies obtained 
via EJE and IS approaches. This indicates that the two methodologies are
consistent and able to reproduce equilibrium properties of the examined 
system. Moreover, for the studied model the structural transitions induced
by pulling can be related to thermodynamical aspects of folding.
\end{abstract}



\maketitle
The properties of the (free) energy landscape can heavily influence the 
dynamical and thermodynamical features of a large class of systems: supercooled liquids, glasses, atomic clusters and biomolecules \cite{wales}.
In particular, the shape of the landscape plays a major role in
determining the folding properties of proteins \cite{funnel}.  
A fruitful approach to the analysis of the landscape relies on the identification
of the local minima of the potential energy, i.e. the ``inherent structures'' (ISs) of
the system \cite{still2}. The investigation of the ISs has lead to the identification of
the structural--arrest temperature in glasses \cite{sastry} and supercooled
liquids \cite{angelani}. More recently, this kind of analysis has been extended to the 
study of proteins \cite{baum,nakagawa}. 

 Mechanical unfolding of single biomolecules represents a powerful technique 
to extract information on their internal structure as well as on
their unfolding and refolding pathways \cite{exp_pap}. 
However, mechanical unfolding of biomolecules is an out-of-equilibrium
process: unfolding events occur on time scales much shorter than the
typical relaxation time of the molecule towards equilibrium.
Nonetheless, by using the equality introduced by Jarzynski \cite{jarzynski}, the 
free energy of mechanically manipulated biomolecules can be recovered as a function 
of an externally controlled parameter \cite{ritort}.

In this Letter, we reconstruct the equilibrium free energy landscape (FEL)
associated to a mesoscopic off-lattice protein model as a function of an
internal coordinate of the system (namely, the end-to-end distance $\zeta$).
At variance with previous studies~\cite{imparato,seifprl,imparato1},  
here we exploit two independent methods: one based on an extended version of 
the Jarzynski equality (EJE) and the other on thermodynamical averages over ISs.
Moreover, the agreement of the results obtained with the two approaches 
indicates that these two methodologies can be fruitfully integrated to
provide complementary information on the protein landscape.
In particular the investigation of the ISs allows us to give an estimate
of the (free) energetic and entropic barriers separating the
native state from the completely stretched configuration. 

The model studied in this paper is a modified version of the 3d off-lattice
model introduced in Ref.~\cite{honey} and successively 
generalized to include a harmonic interaction between
next-neighbouring beads instead of rigid bonds~\cite{berry,veit}.
The model consists of a chain of $46$ point-like monomers 
mimicking the residues of a polypeptidic chain, where each residue
is of one of the three types: hydrophobic ($B$),
polar ($P$) and neutral ($N$) ones. 

The residues within the protein interact via an off-lattice coarse-grained potential 
composed of four terms: a stiff nearest-neighbour harmonic potential intended to maintain the bond
distance almost constant; a three-body bending interaction associated to 
the bond angles; a four-body interaction mimicking the
torsion effects; and a long--range Lennard-Jones potential reproducing in an effective way the
solvent mediated interactions between pairs of residues non covalently bonded \cite{units}.
The $46$-mer sequence $B_9N_3(PB)_4  N_3 B_9  N_3 (PB)_5P$, which exhibits a four stranded $\beta$-barrel
Native Configuration (NC), is here analyzed with the same potential and parameter set reported in Ref. \cite{veit},
but we neglect any diversity among the hydrophobic residues. This sequence has been previously 
studied, for different choices of the potential parameters, in the context of
spontaneous folding \cite{honey,guo,veit,berry,kim} as well
as of mechanical unfolding and refolding \cite{cinpull,lacks}. 
The NC is stabilized by the attractive hydrophobic interactions among the $B$ residues, 
in particular
the first and third $B_9$ strands, forming the core of the NC,
are parallel to each other and anti-parallel to the second and fourth strand, 
namely, $(PB)_4$ and $(PB)_5P$. The latter strands are exposed towards the
exterior due to the presence of polar residues.

The main thermodynamic features can be summarized with reference to 
three different transition temperatures \cite{wales,baum,tlp}: 
the $\theta$-temperature $T_\theta$ discriminating between phases
dominated by random-coil configurations rather than collapsed ones;
the folding temperature $T_f$, below
which the protein stays predominantly in the native valley;
and the glassy temperature $T_g$ indicating the freezing of
large conformational rearrangements \cite{nakagawa}.
Following the procedures reported in Ref. \cite{tlp},
we have determined these temperatures and obtained
$T_\theta = 0.65(1)$, $T_f = 0.28(1)$, and $T_g=0.12(2)$.

 In order to mimic the mechanical pulling of the protein attached to an AFM cantilever,
or trapped in optical tweezers, one extremum of the chain was kept fixed,
and the last bead was attached to a pulling device with a spring of elastic constant $\kappa$. 
The external force is applied at time $t=0$ by moving the device along a fixed direction 
with a constant velocity protocol $z(t)= z(0) + v_p t$.
The protein is initially rotated to have the first and last bead
aligned along the pulling direction, therefore 
the external potential reads $U_{z(t)} (\zeta) =\kappa (z(t) - \zeta)^2/2$. Moreover, to 
reproduce the experimental conditions, the thermalization procedure consists of two steps:
a first stage when the protein evolves freely starting from the NC,
followed by a second one in presence of the pulling apparatus.
The resulting configuration is then used as the starting state 
at $t=0$ for the forced unfolding performed at constant temperature via a 
low friction Langevin dynamics \cite{luccfut}.

Following Ref. \cite{imparato}, we briefly review 
how to reconstruct the equilibrium FEL
as a function of the collective coordinate $\zeta$ starting from 
out-of-equilibrium measurements. Let the system (unperturbed) Hamiltonian 
$H_0(x)$ be a function of the positions and momenta of the residues $x=\{\bm r_i,\bm p_i\}$, 
the free energy of the {\it constrained} ensemble, characterized by a given value $\zeta$
of the macroscopic observable $\zeta(x)$, reads 
$\beta f(\zeta)=- \ln\int \D x\,\delta(\zeta-\zeta(x))\,\E^{-\beta H_0(x)}$.
The system is driven out-of-equilibrium by the external potential, 
$U_{z(t)} (\zeta)$, and the work {\it done} on
the system by the external force associated to $U_{z(t)} (\zeta)$
is $W_t=\int_0^t \D \tau\;v_p \;\kappa\; (z(\tau)-\zeta(x(\tau)))$.
Due to thermal fluctuations the trajectory $x(t)$ followed by the system, and 
therefore $W_t$, varies between one realization of the manipulation process and the other.
In Ref.~\cite{HumSza} an extended version of the Jarzynski equality 
relate $f(\zeta)$ to the work done on the system, for arbitrary external potential. 
Such a relation reads 
\begin{equation}
  \average{\delta(\zeta-\zeta(x))\E^{-\beta W}}_t
  =\E^{-\beta \left(f(\zeta)+ U_{z(t)}(\zeta)\right)}/Z_0,
\label{sample}
\end{equation}
where $Z_0=\int \D x \exp[-\beta H_0(x)]$ and the average $\average{\cdot}_t$ is
performed over different trajectories with fixed time-length $t$.
Technical details for the optimal sampling of the lhs of eq.~(\ref{sample})
are discussed in Refs.~\cite{seifprl,imparato}.
\begin{figure}[t]
\includegraphics[draft=false,clip=true,height=0.25\textwidth, width=0.40\textwidth]{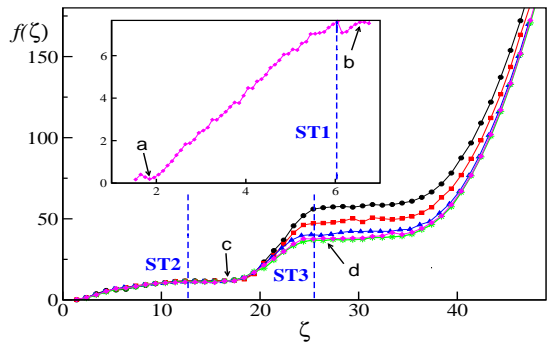}
\caption{(Color online) Free energy profile $f$ as a function of the end-to-end distance $\zeta$, obtained by eq.~(\ref{sample}) for various pulling velocities: from top to bottom
$v_p=5 \times 10^{-2}$, $1 \times 10^{-2}$, $5 \times 10^{-3}$,  $5 \times 10^{-4}$
and $2 \times 10^{-4}$. In the inset, an enlargement of the curve 
for $v_p=5 \times 10^{-4}$ at low $\zeta$ is reported. 
Each curve have been obtained by averaging over $160-240$ repetitions of the same 
pulling protocol at $T=0.3$. The letters indicate the
value of $f(\zeta)$ for the configurations reported in fig.~\ref{fig_conf}
and the (blue) dashed lines the location of the STs.
}
\label{fig_vp}
\end{figure}

\begin{figure}[t]
\includegraphics[draft=false,clip=true,height=0.25\textwidth,width=0.40\textwidth]{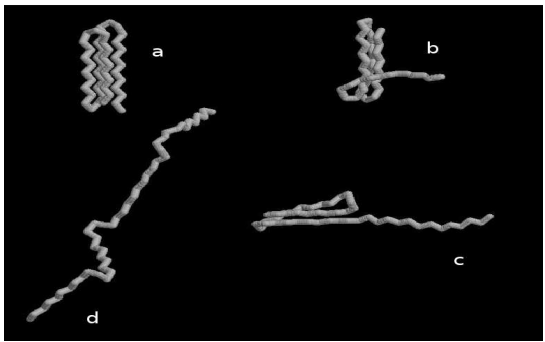}
\caption{Pulled configurations at $T=0.3$: the NC (a) has $\zeta_0 \sim 1.9$; 
the others are characterized by $\zeta=6.8$ (b), $\zeta=16.8$ (c), and $\zeta=27.1$ (d).
}
\label{fig_conf}
\end{figure}

 As shown in fig.~\ref{fig_vp}, the estimated FEL
collapses into an asymptotic curve as the pulling velocity decreases
in agreement with the results reported in~\cite{imparato,imparato1}. 
Let us now discuss, by referring to fig.~\ref{fig_vp} the structural transitions (STs) induced
by the pulling. As shown in the inset, the asymptotic $f(\zeta)$ profile exhibits a 
clear minimum in correspondence of the end-to-end distance of the 
NC (namely, $\zeta_0 \sim 1.9$). Moreover, up to $\zeta \sim 6$, the protein remains in 
native-like configurations characterized by a $\beta$-barrel made up of 4 strands, while the 
escape from the native valley is signaled by the small dip at $\zeta \sim 6$ and it is
indicated in the inset of fig. \ref{fig_vp} as ST1.
This ST has been recently analyzed in \cite{lacks}
in terms of the potential energy of ISs. For higher $\zeta$ the configurations are
characterized by an almost intact core (made of 3 strands) plus a stretched 
tail corresponding to the pulled fourth strand (see (b) and (c) in fig. \ref{fig_conf}).
The second ST amounts to pull the strand $(PB)_5P$ out of the barrel.
In order to do this, it is necessary to break
22 hydrophobic links \cite{links}, amounting to an energy cost $\sim 21$. The corresponding free energy 
barrier height is instead quite lower ($\sim 11$, as estimated from fig.~\ref{fig_vp}). Since 
the potential energy barrier is essentially due to the hydrophobic interactions this
implies that a non negligible entropic cost is associated to ST2.
Instead, in the range $13 < \zeta < 18.5$ the curve $f(\zeta)$ appears as essentially flat,
thus indicating that almost no work is needed to completely stretch the tail once detached from 
the barrel.  The pulling of the third strand (that is part of the core of the NC) leads to a definitive
destabilization of the $\beta$-barrel and to the breakdown of the
remaining 36 BB-links with an energetic cost $\sim 35$. 
A finite entropic barrier should be associated also to
this final stage of the unfolding (termed ST3), because the energy increase due to the hydrophobic terms is 
much higher than the free energy barrier ($\sim 26$, see ST3 in fig.~\ref{fig_vp}).
The second plateau in $f(\zeta)$ corresponds to protein structures made up of a single strand 
(similar to (d) in fig. \ref{fig_conf}). The final quadratic rise of $f(\zeta)$
for $\zeta \geq 36$ is associated to the stretching of bond angles and distances
beyond their equilibrium values.

\begin{figure}[t]
\includegraphics[draft=false,clip=true,height=0.25\textwidth, width=0.4\textwidth]{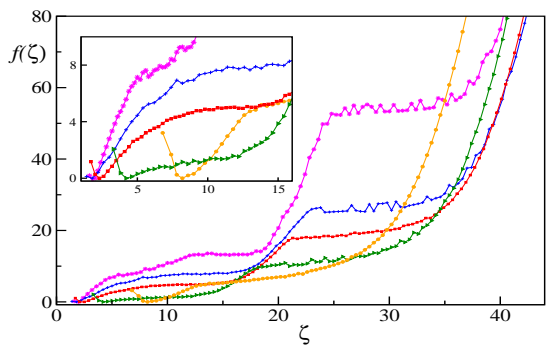}
\caption{(Color online) Free energy profile $f(\zeta)$ as obtained by eq.~(\ref{sample}) for various temperatures:
$T=0.2$ (magenta stars), $0.4$ (blue plus), $0.5$ (red squares), $0.6$ (green
triangles) and $0.7$ (orange circles). In the inset an enlargement is reported
at small $\zeta$. The data refer to $v_p=5 \times 10^{-4}$.}
\label{fig1}
\end{figure}

As shown in fig. \ref{fig1}, the FEL is strongly
affected by temperature variations. In particular, for temperatures
around $T_f$ one still observes a clear minimum around $\zeta_0$ and 
a FEL resembling the one found for $T=0.3$.
A native-like minimum is still observable for
$T =0.5 < T_\theta$, however its position $\zeta > \zeta_0$ indicates that the NC is
no longer the most favourite configuration. Furthermore the dip 
around $\zeta \sim 6 - 7$ disappears and the heights of the two other barriers reduce.
By approaching $T_\theta$ the minimum broadens noticeably and the 
first barrier almost disappears, thus suggesting that 4 stranded $\beta$-barrel 
configurations coexist with partially unfolded ones. Above $T_\theta$  only 
one barrier remains and the absolute minimum is now associated to extended conformations
similar to type (b) or (c) with some residual barrel structure.

Let us now introduce the reconstruction of the free energy in terms of the
inherent states (ISs). ISs  correspond to local minima of the potential energy, 
in particular the phase space visited by the protein during its
dynamical evolution can be decomposed in disjoint attraction basins,
each corresponding to a specific IS \cite{still2,wales}. 
In this context, the free energy can be expressed as a sum over
the basins of attraction:
\begin{equation}
{\rm e}^{-\beta f_{IS}} = \sum_a {\rm e}^{-\beta( V_a + R_a) } \simeq
\sum_a {\rm e}^{-\beta V_a } \prod_{j=1}^{3N-6} (T/\omega_a^j)
\label{zeta} 
\end{equation}
where $a$ labels distinct IS and $V_a$ (resp. $R_a$) is the corresponding potential 
(resp. vibrational free) energy. $R_a$ represents an entropic contribution 
due to the fluctuations around the considered minimum and is analytically estimated 
by assuming a harmonic basin of attraction in terms of the $3N -6$ non zero frequencies 
$\{ \omega_a^j \}$ of the vibrational modes~\cite{wales}.
The harmonic approximation works reasonably well up to $T \sim T_\theta$,
as we have verified by a direct evaluation of the occupation probabilities of the various 
basins~\cite{luccfut}.
We have built up two data banks of ISs:
the thermal data bank (TDB) obtained by performing equilibrium canonical simulations and the 
pulling data bank (PDB) by mechanically unfolding the protein \cite{IS}.

\begin{figure}[t]
\includegraphics[draft=false,clip=true,height=0.25\textwidth, width=0.40\textwidth]{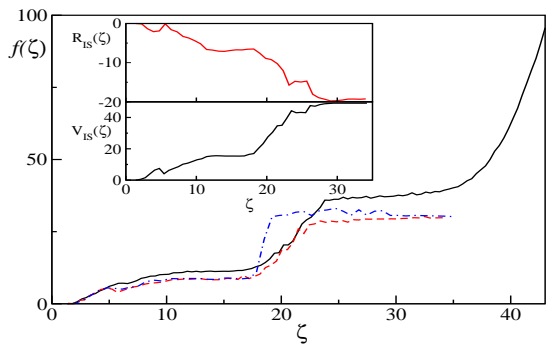}
\caption{(Color online) Free energy profiles $f(\zeta)$ and $f_{IS}(\zeta)$ 
as a function of the elongation $\zeta$ for $T=0.3$. The black solid line refers to the 
reconstruction in terms of the EJE, while the red dashed one corresponds to $f_{IS}$ 
for a set of pulling experiments with $v_p = 2 \times 10^{-4}$. 
The blue dot-dashed line is the $f_{IS}(\zeta)$ obtained in terms of the ISs of the TDB.
In the insets are reported the reconstructed $V_{IS}(\zeta)$ (lower
panel) and $R_{IS}(\zeta)$ (upper panel) by employing ISs in the PDB. 
}
\label{fig2}
\end{figure}

In order to estimate the FEL $f_{IS}(\zeta)$
as a function of the variable $\zeta$ characterizing different ISs, 
the sum in (\ref{zeta}) should be restricted to ISs with an end-to-end distance within 
a narrow interval $[\zeta; \zeta +d \zeta]$ \cite{nakagawa}. As shown in fig. \ref{fig2},
the comparison between $f_{IS}(\zeta)$ and the $f(\zeta)$ obtained via the EJE reconstruction 
in proximity of $T_f$ reveals an almost complete coincidence up to $\zeta \sim 5$, while for larger $\zeta$,
$f_{IS}(\zeta)$ slightly underestimates the free energy. This disagreement is mainly due to the
fact that the IS analysis is based only on minima of the potential, while saddles are completely 
neglected. The further comparison between the IS
reconstruction obtained via the TDB and PDB clearly indicates that the out-of-equilibrium process 
consisting in stretching the protein is more efficient to investigate the FEL, 
since a much smaller number of ISs are needed to well reconstruct it (at least up to $\zeta \sim 17$).

The last stage of the unfolding, reveals a difference among the two $f_{IS}$:
the TDB FEL is steeper with respect to the PDB one, thus suggesting that the protein can reach
lower energy states with large $\zeta$ during mechanical unfolding, states that have a low
probability to be visited during the dynamics at thermal equilibrium. However the value of the barrier to overcome and that of the 
final plateau are essentially the same. The IS conformation with the maximal end-to-end
distance is the all {\it trans}-configuration \cite{trans} corresponding to $\zeta_{trans}=35.70$,
therefore the IS approach does not allow to evaluate the FEL for $\zeta > \zeta_{trans}$.
However, the IS analysis provides us an estimate of the profiles of the
potential and vibrational free energies $V_{IS}(\zeta)$ and $R_{IS}(\zeta)$, respectively. 
From the latter quantity, the entropic costs associated to the unfolding stages can be estimated.
As shown in the inset of fig.~\ref{fig2} for $T=0.3$ the unfolding stages
previously described correspond to clear "entropic" barriers. In particular,
in order to stretch the protein from the NC to the all {\it trans}-configuration
the decrease of $R_{IS}(\zeta)$ is $\sim 19$, in agreement with the previous estimate 
obtained by considering the EJE reconstruction of the FEL.

Finally, one can try to put in correspondence the three unfolding stages previously 
discussed with thermodynamical aspects of the protein folding. In particular,
by considering the energy profile $V_{IS}(\zeta)$, an energy barrier $\Delta V_{IS}$  
and a typical transition temperature $T_t = (2 \Delta V_{IS})/(3N)$,
can be associated to each of the STs. The first transition ST1 corresponds to a barrier 
$\Delta V_{IS} \simeq 8$ and therefore to $T_t \simeq 0.12$, that, within error bars, coincide with $T_g$. 
For the ST2 transition the barrier to overcome is $\Delta V_{IS} \simeq 16$ and this is associated 
to a temperature $T_t \simeq 0.23$ (slightly below $T_f$). The energetic cost to completely stretch 
the protein is $\simeq 49.7$ with a transition temperature $T_t \simeq 0.72$, that is not too far from 
the $\theta$-temperature.
At least for this specific model, our results indicate that the observed STs
induced by pulling can be put in direct relationship with the thermal transitions usually identified 
for the folding/unfolding process.

We can conclude by noticing that the information obtained by the equilibrium FEL 
both with the EJE and the IS methodologies are consistent  
and give substantiated hints about the thermal unfolding.
However, we want to point out that these two methods
are somehow complementary. On the one hand, with the EJE 
approach all the coordinates are projected onto a collective one,
the contribution of the microscopic configurations being averaged out.
On the other hand, the IS analysis appears more suitable to study the 
microscopic details of the configuration space of complex systems such as 
proteins, once the main basins have been identified by using the former approach.

\acknowledgments
Useful discussions with the members of the CSDC in Firenze and L. Peliti are acknowledged,
as well as partial support by the European Contract No. 12835 - EMBIO.



\begin{thebibliography}{99}
\bibitem{units} All quantities are here expressed in adimensional units 
for a comparison with physical units see \cite{veit}.
\bibitem{IS} In order to find the different ISs the equilibrium 
(resp. out-of-equilibrium) Langevin trajectory is sampled at constant time intervals 
$\delta t = 5$ (resp. at constant elongation increments 
$\delta \zeta = 0.1$) to pinpoint a series 
of configurations, which afterwards are relaxed via a steepest descent dynamics.
For mechanical unfolding, the protein is unblocked and the pulling apparatus removed
before the relaxation stage. The TDB contains $\sim 600,000$ distinct ISs collected 
via equilibrium simulations at various temperatures in the range $[0.3;2.0]$. The PDB
contains $5,000 - 50,000$ ISs for each examined temperature.
\bibitem{links} We assume that a hydrophobic contact is formed between two B-residues
if their distance is smaller than $1.25$.
\end{thebibliography}
\end{document}